\documentclass[pra,amssymb,letterpaper,twocolumn,nofootinbib]{revtex4}

\usepackage{amsmath}
\usepackage{epsfig}
\usepackage{amsfonts}
\usepackage{graphicx}

\newenvironment{figurehere}
   {\def\@captype{figure}}
   {}

\def\>{\rangle}
\def\<{\langle}

\def\be{\begin{equation}}
\def\ee{\end{equation}}

\begin{document}
\title{Quantum searching a classical database\\
(or how we learned to stop worrying and love the bomb)}

\author{Terry Rudolph}
\email{rudolpht@bell-labs.com}
\author{Dr.(Strange)Lov Grover}
\affiliation{Bell Labs, 600-700 Mountain Ave., Murray Hill, NJ 07974, U.S.A.}

\date{\today}

\begin{abstract}
We show how to perform a quantum search for a classical object,
specifically for a classical object which performs no coherent
evolution on the quantum computer being used for the search. We
do so by using interaction free measurement as a subroutine in a
quantum search algorithm. In addition to providing a simple
example of how non-unitary processes which approximate unitary
ones can be useful in a quantum algorithm, our procedure requires
only one photon regardless of the size of the database, thereby
establishing an upper bound on the amount of energy required to
search an arbitrarily large database. Alternatively, our result
can be interpreted as showing how to perform an interaction free
measurement with a single photon on an arbitrarily large number of
possible bomb positions simultaneously. We also provide a simple
example demonstrating that in terms of the number of database
queries, the procedure outlined here can outperform the best
classical one.
\end{abstract}


\maketitle

{\sl  The nefarious Dr.~Strangelove has created a doomsday bomb so
sensitive that if it absorbs a \emph{single} photon it will
explode. More distressing than this, even more distressing in fact
than Strangelove's strange accent, is that you are his prisoner.
In his typical dastardly manner, he toys with you by informing you
that the bomb is in one of a large number $N$ of boxes, and he
will release you and disarm the bomb only if you can determine
which one. However, in the clich\'ed manner of all evil
overlords, Strangelove decides to grant you (and the rest of the
world for that matter) some chance at life, by allowing you
standard optical elements and photodetectors to aid in your
challenge.

When you fail to exhibit the abject desperation upon which
Strangelove thrives, in fact you even look a little smug,
Strangelove becomes suspicious. He calls in his evil assistant
Petraskudo, who is, as all evil assistants should be, a rather
good physicist.  Petraskudo informs Strangelove that in fact there
\emph{is} a way, using `interaction-free' measurement
\cite{ifm} for you to locate the box holding the bomb. However,
she suggests that if Strangelove changes the rules, as is the
clear perogative of an evil overlord, in such a way as to let you
use only \emph{one} photon, then you will only be able to search a
single box. Unfortunately, Petraskudo is not as smart as she
thinks she is; she should have abandoned reading refereed
journals and started reading only the quant-ph archive -- on
which she would have found a long precedent for quantum
information theorists trying to rescue the imprisoned\cite{me},
and in particular she'd have found the paper \cite{terry} which
explains how to determine the box holding the bomb using only one
photon...}

The quantum search algorithm \cite{lov} was originally phrased in
terms of searching an unsorted database for a marked item. This
was unfortunate; it allowed particularly polemic people such as
Charlie Bennett, to  argue that such a database would have to be
a specially constructed ``quantum'' database, and could not be a
regular classical database. The question as to exactly what a
classical database \emph{is} does not seem to have been
addressed. Accordingly, as with so many things in quantum
information theory, we choose here to make up our own definition
-- a definition which of course conveniently coincides with a
problem we can solve. We should point out that the original paper
on quantum searching in fact contained a completely cryptic
comment about interaction free measurement (IFM) and its potential
utilization as part of a quantum search. We therefore pretend
here that this was not recalled with hindsight, but rather formed
part of the motivation for this work.

We will imagine that some classical object is used to mark a
particular one of $N$ items, the other items have no such
object\footnotemark[1]\footnotetext[1]{Since empty space is a
particularly good quantum channel for photons (although not for
two-level atoms), one might argue that this is not a completely
classical database. We have no reply, except to say that all
information (quantum or classical if your personal philosophy
insists on a distinction), is encoded in/carried by physical
systems and is thus describable within quantum theory. However
the abstract notion of `a classical object' is \emph{not}
axiomatized within quantum theory, and thus defining a classical
database in a manner unobjectionable to all will be difficult.
Moreover, since this problem is ultimately responsible for the
employment of truckloads of physicists and philosophers working
in quantum foundations, we suspect that concerns over job
security has prevented most of them making any sort of sensible
effort on this problem.}. Obviously, if the classical object is a
specially chosen phaseshifter, then a quantum search for the
phaseshifter is simple. However, this is generally considered
``not cricket'', as an (East) Indian might say. We therefore
assume the object provides no potential coherent evolution for a
photon; it is some form of incoherent scatterer (or doomsday
bomb). By using interaction free measurement as a subroutine in a
quantum search algorithm we will show how the marked item can be
located using only a single photon, thereby setting a fairly
small upper bound on the total amount of energy required to
search an arbitrarily large database. Our algorithm also provides
a simple demonstration of how a series of non-unitary operations,
which asymptotically approach a unitary one, can be used as part
of a coherent quantum computation. Finally, we will give a simple
example which demonstrates that in terms of the number of database
``queries'' the algorithm presented here can outperform the best
classical one.
\begin{figurehere}
\center{\epsfig{file=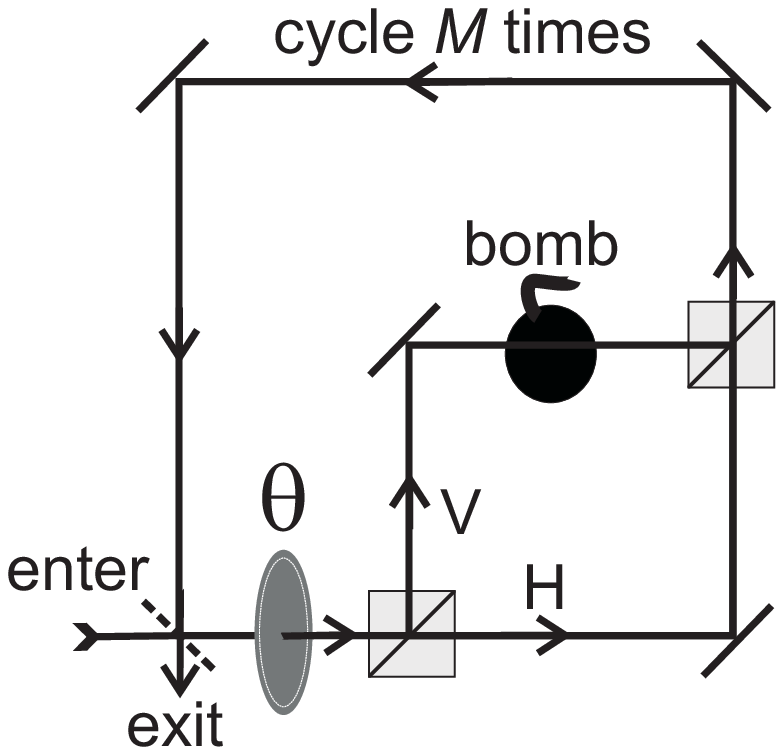,width=50mm,clip=}}
\\*
{\bf Fig.1.}{\small `Interaction free' detection of a single
bomb.}
\end{figurehere}
\vskip 0.3cm

We begin by briefly reviewing the standard scheme for
IFM\footnotemark[2], as shown in
Fig.1.\footnotetext[2]{Interaction free measurement is also known
by the more politically correct term `quantum interrogation'.
This paper makes no pretentions towards political correctness.} A
horizontally polarized photon (in state $|H\>$) enters an
interferometer through a switchable mirror, and is rotated an
angle $\theta$ by a polarization rotator (described by the matrix
${\cos\theta\;-\sin\theta \choose
\sin\theta\;\;\cos\theta}$ in the $|H\>,|V\>$ basis).
It then passes to a polarizing beam splitter(PBS), which
transmits horizontally polarized light and reflects vertically
polarized light. If there is no bomb in the vertical arm of the
polarization interferometer then the photon emerges from the
second PBS in the state
$|\theta\>\equiv\cos\theta|H\>+\sin\theta|V\>$. If there \emph{is}
a bomb, then either the bomb explodes (with probability
$\sin^2\theta$), or the photon is collapsed back to the state
$|H\>$. The photon is fed back through the whole system $M$
times, after which it exits the circuit via the switchable
mirror. If there is no bomb then it exits in the state
$|M\theta\>$, if there is a bomb, then with probability
$\cos^{2M}\theta$ it exits in the state $|H\>$ and with
probability $1-\cos^{2M}\theta$ the bomb explodes. If we choose
$\theta=\pi/2M$, then the probability of \emph{not} exploding the bomb
approaches 1 for large $M$; moreover the photon emerges in the
state $|H\>$ or $|V\>$ according to whether or not a bomb is
present, and these two states are orthogonal and thus
distinguishable.

We note that it is irrelevant \emph{where} in the vertical arm of
the interferometer the bomb is located, in effect the photon is
``sniffing out'' a large number of spacetime points. Thus we
intuitively expect that there should be some way of modifying the
above procedure to provide information on an arbitrarily large
number of potential bomb positions. To do so we first change the
above procedure to choose $\theta=\pi/M$. Thus, after $M$ cycles
the state will be $\pm|H\>$ according to whether or not a bomb is
present. That is, the presence of the bomb introduces a $\pi$
phase shift to the probing photon.
\vspace{-0.5truecm}
\begin{figurehere}
\center{\epsfig{file=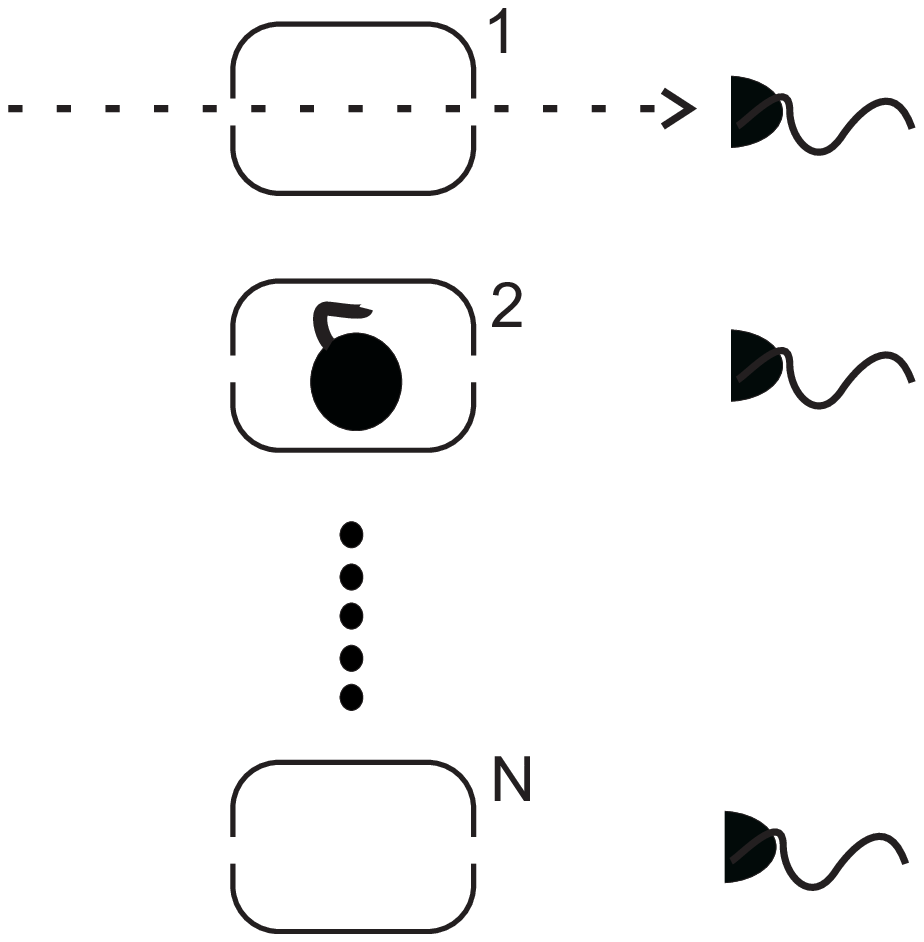,width=45mm,clip=}}
\\*
{\bf Fig.2.}{\small  A series of $N$ boxes, one of which contains
the bomb.}
\end{figurehere}
\vskip 0.2cm
We assume the bomb is known to be located in one of an arbitrarily
large number $N$ of boxes, as depicted in Fig.2. An IFM device
similar to the one in Fig.1. is built around each box, with
$\theta$ chosen as discussed above. A horizontally polarized
photon is placed in a superposition of states, such that it has
equal amplitude for being in each of the spatial modes 1 through
$N$. This state can be written
$|\psi^{(0)}\>=\frac{1}{\sqrt{N}}\sum_{i=1}^N|H_i\>$. The photon
then goes through the $N$ different IFM setups simultaneously,
the $M$ cycles of which we will refer to as ``small cycles''.
Assuming the bomb does not explode, which occurs with probability
$P_{\neg x}^{(1)}=1-\frac{1-\cos^{2M}\theta}{N}$, the photon
emerges after the $M$ small cycles in the state
$|\psi^{(0)'}\>=\left(\sum_{i\neq
t}|H_i\>-\cos^M\theta|H_t\>\right)/\sqrt{P_{\neg x}^{(1)}}$, where
we label the `target' mode holding the bomb by $t$ and drop an
overall phase of $\pi$. We arrange for the photon to pass through
an array of beamsplitters designed
\cite{reck} to perform the ``inversion about average'' operation
\cite{lov}. This completes the first ``large cycle'', the state at
the end of which is given by $|\psi^{(1)}\>=\sum_{i\neq
t}\alpha^{(1)}|H_i\>+\tau^{(1)}|H_t\>$. If we go through $k$
large cycles (without exploding the bomb), then it can be shown
that the state of the photon is $|\psi^{(k)}\>=\sum_{i\neq
t}\alpha^{(k)}|H_i\>+\tau^{(k)}|H_t\>$, where
$\alpha^{(k)}=\frac{\tilde{\alpha}^{(k)}}{\sqrt{(N-1)\tilde{\alpha}^{(k)2}+\tilde{\tau}^{(k)2}}}$,
$\tau^{(k)}=\frac{\tilde{\tau}^{(k)}}{\sqrt{(N-1)\tilde{\alpha}^{(k)2}+\tilde{\tau}^{(k)2}}}$,
and
\begin{eqnarray}
\tilde{\alpha}^{(k)}&=&-\cos^M\theta\sin
k\phi+\sqrt{\cos^{M}\theta}\sin(k+1)\phi \nonumber\\
\tilde{\tau}^{(k)}&=&\sin k\phi+\sqrt{\cos^{{M}}\theta}\sin(k+1)\phi.
\end{eqnarray}
Here $\phi$ is defined by
$$\cos\phi=\frac{\left(1-\frac{2}{N}\right)(1+\cos^M\theta)}{2\sqrt{\cos^{M}\theta}},$$
and we recall that $\alpha^{(0)}=\tau^{(0)}=1/\sqrt{N}$ and
$\theta=\pi/M$. The cumulative probability that we get to large
cycle $k$ without the bomb exploding is given by
\begin{eqnarray}
{\cal P}(k)&=&\prod_{i=1}^k P_{\neg x}^{(i)}=\prod_{i=1}^k
\left(1-\tau^{(i-1)2}(1-\cos^{2M}\theta)\right)\nonumber\\
&\ge& \prod_{i=1}^k \cos^{2M}\theta=\cos^{2kM}\theta.
\end{eqnarray}

In the case of large $N$, it is not difficult to show from (1) and
(2) that by choosing $k=O(\sqrt{N})$ and $M$ to be asymptotically
greater than $k$, the probability of the photon ending up in mode
$t$ without exploding the bomb goes to 1. However, the evolution
described by the equations (1),(2) is quite complicated in
general.
\vspace{-0.8cm}
\begin{figurehere}
\center{\epsfig{file=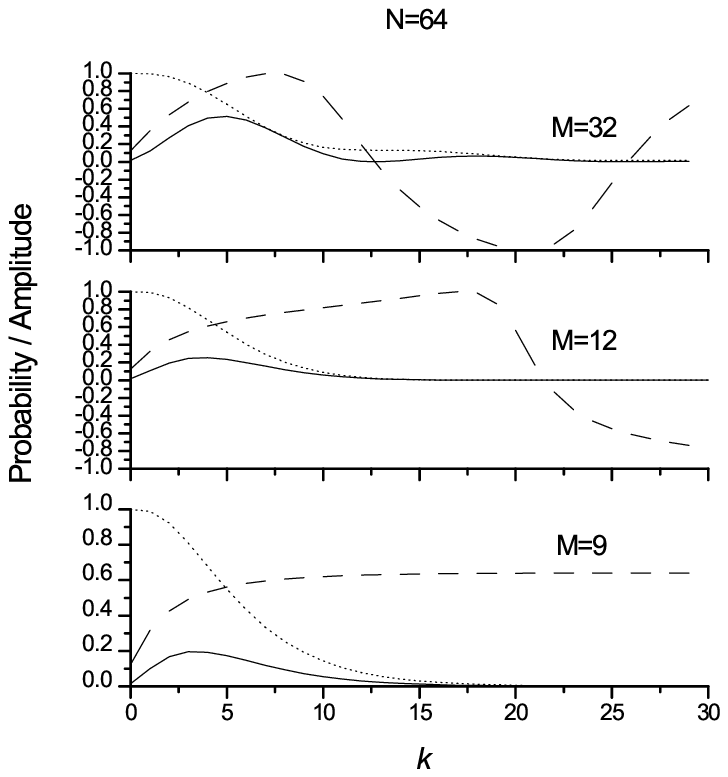,width=90mm,clip=}}
\\*
{\bf Fig.3.}{\small Graphs of $\tau^{(k)}$ (dashed line), ${\cal
P}(k)$ (dotted line) and ${\cal P}(k)|\tau^{(k)}|^2$ (solid line)
versus $k$, for $N=64$ and $M=9,12,32$.}
\end{figurehere}

In Fig.3. we show how $\tau^{(k)}$ (dashed line) and ${\cal
P}(k)$ (dotted line) vary with $k$, along with the overall
probability of successful detection ${\cal P}(k)|\tau^{(k)}|^2$
(solid line). For large values of $M$ the evolution approaches
that of the standard quantum search. However for small $M$ the
evolution can be such that the amplitude for the target state
saturates and does not oscillate -- this corresponds to the point
where the parameter $\phi$ changes from being imaginary to real.
In Fig.3. this transition between the saturation and oscillation
occurs for $M$ between 9 and 12.

We conclude by making some remarks about interpreting this
algorithm as a search of a classical database. We presume we need
not search the database in fear of our lives, i.e. that the
classical item marking a certain box is not explosive. After
$k$ large cycle iterations, the total number of database queries
will be
$kM$.  We have not, so far,
been able to establish the extent to which this algorithm may be
used to out perform a classical one. That it
\emph{can} somewhat do so is implied by the following examples.

Let us take $N=4$ and examine the case when we make only 1 query
to the database. Classically, our probability of identifying the
marked item in our
 $1/4$ or $25\%$. For a regular quantum search it would be $100\%$.
If instead we run an algorithm such as the one presented here,
then the probability of successfully identifying the marked item
is $56.25\%$. If instead we take $N=15$ and consider the case when
we make only 3 queries to the database, then classically our
probability of obtaining the marked item in our three queries is
$3/15$ or $20\%$. A regular quantum search would find the marked
item with $93.5\%$ probability of success. However, if we run the
algorithm presented here with
$M=3$ and $k=1$, we find that the probability of successfully identifying
the marked item is now  about $26\%$, still marginally better than
the classical case.

\begin{acknowledgments}
 This research was supported by the NSA \& ARO under contract No.
DAAG55-98-C-0040.
\end{acknowledgments}

\end{document}